\documentclass[aps,showpacs,prl,twocolumn,superscriptaddress,apsrev,english]{revtex4-2}
\usepackage{multirow}
\usepackage{color}
\usepackage{amsfonts}
\usepackage{amsmath}
\usepackage{mathrsfs}
\usepackage{amssymb}
\usepackage{wasysym}
\usepackage{graphicx}
\usepackage{paralist}
\usepackage{indentfirst}
\usepackage{hyperref}
\usepackage{ulem}
\usepackage{soul}
\usepackage{cancel}
\usepackage{CJK}
\usepackage{float}
\usepackage{subfigure}
\usepackage{braket}
\usepackage{hyperref}

\hypersetup{hypertex=true,
	colorlinks=true,
	linkcolor=blue,
	anchorcolor=blue,
	citecolor=blue,
	urlcolor=blue}

\begin{document}
	
\title{Rabi Spectroscopy of Super-Bloch Oscillations in Optical Lattice Clock}%

\author{Sheng-Xian Xiao}
\affiliation{Chongqing Key Laboratory for Strongly Coupled Physics, Chongqing University, Chongqing, 401331, China}
\affiliation{Center of Modern Physics, Institute for Smart City of Chongqing University in Liyang, Liyang 213300, China}
\author{Ying Liang}
\affiliation{Chongqing Key Laboratory for Strongly Coupled Physics, Chongqing University, Chongqing, 401331, China}

\author{Ya Zhang}
\affiliation{School of Instrumentation Science and Engineering, Harbin Institute of Technology, China}

\author{Tao Wang}
\thanks{corresponding author: tauwaang@cqu.edu.cn}
\affiliation{Chongqing Key Laboratory for Strongly Coupled Physics, Chongqing University, Chongqing, 401331, China}
\affiliation{Center of Modern Physics, Institute for Smart City of Chongqing University in Liyang, Liyang 213300, China}

\begin{abstract}
Super-Bloch oscillations(SBOs) is giant Bloch oscillations (BOs) when applying both static and periodically driving force to free atoms in lattice at the condition that Bloch oscillations are close to integer times of driving frequencies. Rather than observe SBOs in real space, this paper presents a method to observe it using Rabi spectroscopy of Optical lattice clock(OLC). An effective model of OLC with atoms been added both static and time-periodical forces is derived. Based on that, we propose an experimental scheme and give the Rabi spectrum under lab achievable parameters. Utilizing the precision spectroscopy of OLC, force with a large range could be accurately measured by measuring the Period of SBOs. We also gave the best parameter condition of measuring gravity by calculating Fisher information. Our work paves the way to study other exotic dynamics behaviors in Floquet driving OLC. 
\end{abstract}
\maketitle

\section{INTRODUCTION}
Bloch oscillations will occur in real space when a constant force $F$ is given to electrons in a perfect lattice, as predicted around 80 years ago by Bloch and Zener \cite{BO1,BO2}. Using the dispersion relationship, we can figure out the frequency of BOs: $\nu_B=Fd/h$ with lattice constant $d$. Due to requiring long coherent time, BOs were not discovered in superlattices until 1992 \cite{superlattice1,superlattice2}, and subsequently observed in optical lattices \cite{OL}, optical lattice clock \cite{OLC}, superconducting quantum processors \cite{SQB}, etc. BOs are not noly a measurement technique for Floquet-Bloch spectrum \cite{F-B1,F-B2}, but also used for quantum precision measurements, such as measurements of gravitational acceleration \cite{Gravity1,Gravity2}. 

By adding a second time-periodic force, whose frequency is slightly out of tune with $w_B$, the Bloch oscillations' amplitude and period will be hundreds of times bigger \cite{SB0s-ep}. This phenomenon is called the super Bloch oscillations. On the basis of the effective dispersion relation, SBOs can be simply interpreted as Bloch oscillations \cite{SBOs-th1,SBOs-th2}. However, since SBOs have been found, in-situ measurement has become the primary method of observation for SBOs, which significantly reduces the precision of observations and the range of possible applications. Recent advancements in Floquet Engineering's optical lattice clock (FE-OLC) have allowed us to see SBOs spectroscopy in OLC \cite{FE-OLC}, offering yet another method of observing SBOs.  While this is going on, OLC, one of the most accurate experimental platforms for frequency \cite{frequency}, may be able to identify SBOs with much greater precision, which could increase the accuracy of measurements of gravitational acceleration.

Here, we propose an experimental method for observing SBOs in conjunction with FE-OLC and its application for measurements of gravitational acceleration. Additionally, our paper is structured as follows: We derive the theoretical model of OLC in Section \ref{sec2} by applying both time-periodic and constant forces. in Section \ref{sec3} and \ref{sec4}, we show how to prepare atoms whose distribution is momentum-dependent in the FE-OLC and how to realize the Rabi spectrum of SBOs, respectively. In Section \ref{sec5}, we discuss the best experimental scheme for gravitational acceleration measurement in FE-OLC by calculating the Fisher information.

\section{THE MODEL \label{sec2}}
The optical lattice is formed by two counter-propagating lattice laser beams produced by splitting one
lattice laser. The amplitude and frequency of the weak lattice laser can be changed by a voltage variable attenuator and an acousto-optic modulator. The frequency of weak lattice laser can be periodically driven via arbitrary function generator. Therefore, the periodic driving lattice potential can be written as:
\begin{eqnarray}
	U(t)&=&-\frac{U_z}{2}\cos(2k_Lz-\pi\int_{0}^{t}{\Delta\nu(\tau)d\tau})\nonumber\\
	&\;&+\frac{2U_rr^2}{W_0^2}-\eta\frac{2U_rr^2}{W_0^2}\sin^2(k_Lz),\label{eq1}
\end{eqnarray}
where $U_z$ is the potential in the $z$ direction, $U_r$ is the potential in the $r$ direction, $\eta=U_z/U_r$ is the coupling constant, $k_L=2\pi/\lambda_L$ with the lattice laser wavelength $\lambda_L$, $r=\sqrt{x^2+y^2}$ represents the position of the atoms perpendicular to the lattice direction $z$, and $\Delta\nu(t)$ is periodically driving function with frequency $\nu_s$.

The interaction between clock laser with atom after rotating wave approximation(RWA) is:
\begin{eqnarray}
	\hat{H}_{int}^{RWA}=\frac{h\delta}{2}\sigma_z+\frac{h g_0}{2}(e^{ik_pz}\sigma_++e^{-ik_pz}\sigma_-),\label{eq2}
\end{eqnarray}
where $\delta$ is the detuning of the clock laser, $g_0$ is bare Rabi frequency, $k_p=2\pi/\lambda_p$ with  the clock laser wavelength $\lambda_p$ and $h$ is Plank's constant. $\sigma_z$ and $\sigma_{\pm}=\sigma_x\pm i\sigma_y$ are Pauli matrices.

Then, considering the constant force $F_0$(usually provided by gravity in OLC), we can write the total Hamiltonian as:
\begin{eqnarray}
	\hat{H}_{total}(t)=\frac{\textit{\textbf{p}}^2}{2M}+U(t)+\hat{H}_{int}-F_0z,\label{eq3}
\end{eqnarray}
where $\textit{\textbf{p}}$ and $M$ are respectively the momentum and mass of atom. 
For SBOs, there is off-resonance between $F_0$ and $\nu_s$ : $F_0d=(n+\Delta)h\nu_s$ ($n$ is integer and $\Delta\ll1$ ), where $d=\lambda_L/2$ is the lattice constant. 

Eq.(\ref{eq3}) can be re-represented using the creation (annihilation) operators $\hat{c}_{l,\vec{n},\sigma}^\dagger$ ($\hat{c}_{l,\vec{n},\sigma}$), where $\sigma=e,g$ labels the internal state of atom. In the $z$ direction, we choose wannier basis $\left|l,n_z\right\rangle$, where $l$ denotes the site number and $n_z$ is the band index. In the $r$ direction, we choose the harmonic eigen basis $\left|n_x,n_y\right\rangle$ with the eigen-energy $h \nu_r(n_x+n_y+1)$ in which the radial frequency is $\nu_r=\sqrt{\frac{U_r}{M\pi^2W_0^2}}$. Thus, in the co-moving frame in $z$ direction, the total Hamiltonian becomes:
\begin{eqnarray}
	\hat{H}(t)&=&-\sum_{\vec{n},\sigma}\sum_{l,l^{\prime}}J_{l^{\prime}l}^{\vec{n}}\hat{c}_{l^{\prime},\vec{n},\sigma}^\dagger\hat{c}_{l,\vec{n},\sigma}+\hat{H_r}\nonumber\\
	&&-\left(F_0+F(t)\right)d\sum_{l,\vec{n},\sigma}l\hat{c}_{l,\vec{n},\sigma}^\dagger\hat{c}_{l,\vec{n},\sigma}\nonumber\\
	&&+\sum_{l,\vec{n}}[\frac{h(\delta-\frac{v(t)}{c}\nu_p)}{2}\left(\hat{c}_{l,\vec{n},e}^\dagger\hat{c}_{l,\vec{n},e}-\hat{c}_{l,\vec{n},g}^\dagger\hat{c}_{l,\vec{n},g}\right)\nonumber\\
	&&+\frac{h g_{\vec{n}}}{2}\left(e^{il\Phi}\hat{c}_{l,\vec{n},e}^\dagger\hat{c}_{l,\vec{n},g}+h.c\right)] ,\label{eq4}
\end{eqnarray}

where $c$ is the speed of light, $\nu_p$ is the frequency of clock laser and $\hat{H_r}=\sum_{l,\vec{n},\sigma}h \nu_r(n_x+n_y+1)\hat{c}_{l,\vec{n},\sigma}^\dagger\hat{c}_{l,\vec{n},\sigma}$.
The parameters  $g_{\vec{n}}=g_0\left \langle n_z,0 \right |e^{ik_pz}\left|n_z,0\right \rangle$ is the modified Rabi frequency, $\Phi=\pi\lambda_L/\lambda_p$ is the spin-orbit coupling effect and $J_{l^{\prime}l}^{\vec{n}}$ contains two parts: the hopping term $J_{l^{\prime}l}^{n_z}=-\left \langle 0,n_z \right|\frac{\hat{p}_z^2}{2M}-\frac{U_z}{2}\cos(2k_Lz)\left|l-l^{\prime},n_z\right\rangle$ and the coupling term $C_{l^{\prime}l}^{\vec{n}}=\eta\frac{h \nu_r}{2}(n_x+n_y+1)\left \langle 0,n_z \right|\sin^2(k_Lz)\left|l-l^{\prime},n_z\right\rangle$. $F(t)=\frac{M\lambda_L}{2}\frac{d\Delta\nu(t)}{dt}$ and $v(t)=\frac{\lambda_L}{2}\Delta\nu(t)$ are respectively the effective force and velocity with the same period $T_s$ induced in the co-moving frame in $z$ direction.

The presence of constant force $F_0$ seems to prevents us from following the usual steps to deal with the above time-dependent Hamiltonian Eq.(\ref{eq4}) \cite{FE-OLC,Bessel1,Bessel2}, in particular the off-resonance between constant forces and driven. However, if the non-resonant part $\Delta\nu_s$ is smaller than the other parameters, we can retain it as a slow variation. Then, we can get the rotating Hamiltonian $\hat{H}_R^m(t)=\hat{U}_2^\dagger\hat{U}_1^\dagger\left(\hat{H}(t)-i\hbar\frac{\partial}{\partial t}\right)\hat{U}_1\hat{U}_2$ by using two following unitary operators:
\begin{eqnarray}
	\hat{U}_1&=&\exp\left[i\sum_{l,\vec{n},\sigma}X_l(t)\hat{c}_{l,\vec{n},\sigma}^\dagger\hat{c}_{l,\vec{n},\sigma}\right] \label{eq5} \\
	\hat{U}_2&=&\exp\left[iY_m(t)\sum_{l,\vec{n}}\left(\hat{c}_{l,\vec{n},e}^\dagger\hat{c}_{l,\vec{n},e}-\hat{c}_{l,\vec{n},g}^\dagger\hat{c}_{l,\vec{n},g}\right)\right]\label{eq6}
\end{eqnarray}
with
\begin{eqnarray}
	X_l(t)&=&\frac{l}{\hbar}\left[\int_{0}^t(F(\tau)d+nh\nu_s)d\tau+\frac{M\lambda_L^2}{4}\Delta\nu(0)\right], \label{eq7}  \\
	Y_m(t)&=&\int_0^t\frac{\pi v(\tau)\nu_p}{c}d\tau+\pi m\nu_st,\label{eq8}
\end{eqnarray}
and the Floquet sideband index $m$. Considering $\nu_s$ is much bigger than other parameters, the effective Hamiltonian $\hat{H}_{eff}^m=\frac{1}{T_S}\int_0^{T_s}\hat{H}_R^m(t)dt$ can be derived by using Floquet-Magnus expansion:
\begin{eqnarray}
	\hat{H}_{eff}^m&=&-\sum_{\vec{n},\sigma}\sum_{l,l^{\prime}}J_{l^{\prime}l}^{\vec{n}}\mathcal{F}_{l-l^{\prime}}\hat{c}_{l^{\prime},\vec{n},\sigma}^\dagger\hat{c}_{l,\vec{n},\sigma} +\hat{H_r}\nonumber \\
	&&+\sum_{l,\vec{n}} [\frac{h(\delta+m\nu_s)}{2}\left(\hat{c}_{l,\vec{n},e}^\dagger\hat{c}_{l,\vec{n},e}-\hat{c}_{l,\vec{n},g}^\dagger\hat{c}_{l,\vec{n},g}\right)  \nonumber\\
	&&+\frac{h g_{\vec{n}}\mathcal{R}^m}{2}\left(e^{il\Phi}\hat{c}_{l,\vec{n},e}^\dagger\hat{c}_{l,\vec{n},g}+h.c\right)]+\hat{H}_\Delta,\label{eq9}
\end{eqnarray}
where
\begin{eqnarray}
	\hat{H}_\Delta&=&-\Delta h\nu_s\sum_{l,\vec{n},\sigma}l\hat{c}_{l,\vec{n},\sigma}^\dagger\hat{c}_{l,\vec{n},\sigma}\label{eq10}
\end{eqnarray}
is the effective force term that added to the atoms which leads to BOs. The effective coefficients:
\begin{eqnarray}
	\mathcal{F}_{l-l^{\prime}}&=&\frac{1}{T_s}\int_{0}^{T_s}e^{i\theta_{l^{\prime}l}(t)}dt \label{eq11}\\
	\mathcal{R}^m&=&\frac{1}{T_s}\int_{0}^{T_s}e^{i\theta_m(t)}dt\label{eq12}.
\end{eqnarray}
$\theta_{l^{\prime}l}(t)=(l-l^\prime)[ \frac{M\lambda_L^2}{4\hbar}\Delta\nu(t)+2\pi n\nu_st]$ and $\theta_{m}(t)=-(\Phi\int_0^t\Delta\nu(\tau)d\tau+2\pi m\nu_st)$ are the time-dependent phase factors.

\begin{figure}[t]
	\includegraphics[width=0.99\linewidth]{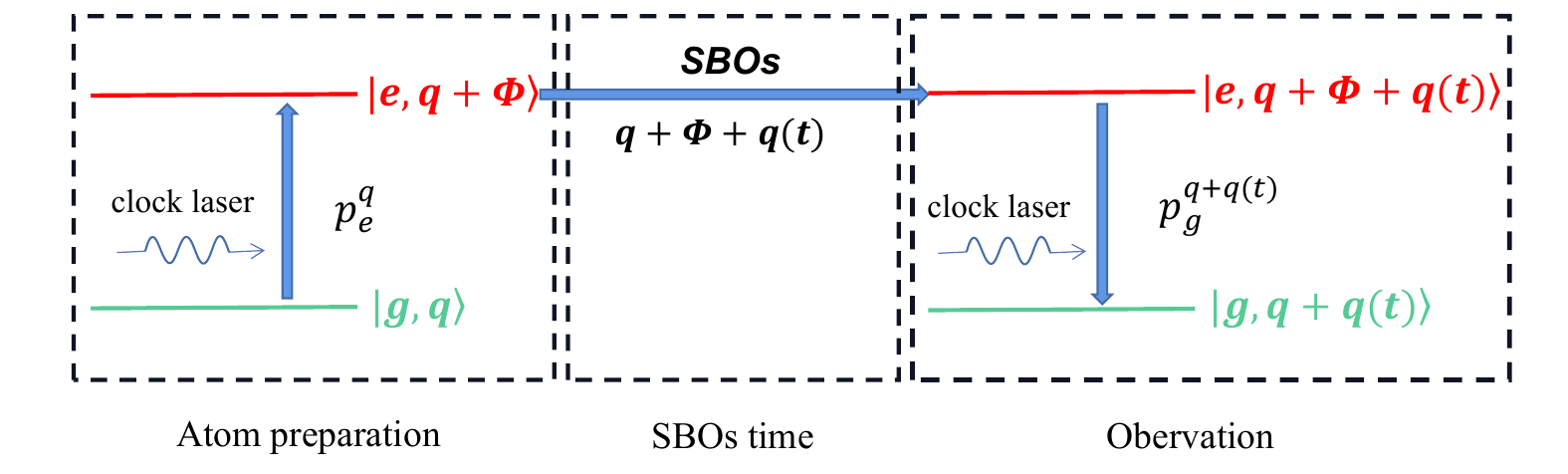}
	\caption{The scheme for observing SBOs in OLC}
	\label{buzhou}
\end{figure}

Based on the analysis above, we propose a scheme for observing SBOs in FE-OLC exhibited in Fig.\ref{buzhou}. In the first step, a $\pi$ pulse clock laser is used to exciting the atoms from $^1S_0$ to $^3P_{0}$ state around a certain momentum $q$, during which $\hat{H}_\Delta$ is ignored because $\Delta\nu_s$ is much smaller than the effective Rabi coupling strength $g_{\vec{n}}\mathcal{R}^m$. In the second step, the clock laser is turned off and atoms are moving in momentum space with a constant velocity due to effective force. The lifetime of $^3P_0$ state is as long as $128$s, which is always longer than BOs time that we setting, thus the spontaneous emission could be ignored.  In the third step, another $\pi$ pulse clock laser is used to bring atoms back to $^1S_0$ state. From the Rabi spectroscopy one can get the information of atoms motion in momentum space. 

Before we go to any detail studies of SBOs, let's briefly summarize the effect of periodically driving the optical lattice according to Eq.(\ref{eq9}). First, it modifies the band dispersion by changing the hopping amplitude; second, it modifies the Rabing coupling strength; third, it reduce the force that added to atoms. The third effect is crucial for measuring the BOs caused by large force by changing it to a small force so that atoms move in the band without Landau-Zener transition.

\section{atoms preparation with momentum-dependent distribution\label{sec3}}  
The effictive time-independent Hamiltonian for $m$th Floquet sideband can be exactly solved in the momentum space if the approximation $\Delta\nu_s\ll g_{\vec{n}}\mathcal{R}^m$ is satisfied and only the nearest neighbor hopping ($|l-l^{\prime}|=1$) is considered. $\hat{H}_{eff}^m$ can be rewritten as:
\begin{eqnarray}
	\hat{H}_{eff}^m&=&\sum_{q,\vec{n},\sigma} E_{\left|\vec{n},q\right>}\hat{c}_{q,\vec{n},\sigma}^{\dagger}\hat{c}_{q,\vec{n},\sigma} \nonumber \\
	&\;&+\sum_{q,\vec{n},\sigma}[\frac{h(\delta+m\nu_s)}{2}\left(\hat{c}_{q,\vec{n},e}^\dagger\hat{c}_{q,\vec{n},e}-\hat{c}_{q,\vec{n},g}^\dagger\hat{c}_{q,\vec{n},g}\right)\nonumber \\
	&\;&+\frac{h g_{\vec{n},m}}{2}\left(\hat{c}_{q+\Phi,\vec{n},e}^\dagger\hat{c}_{q,\vec{n},g}+h.c\right)]\label{eq13}
\end{eqnarray}
with effective Rabi frequency $g_{\vec{n},m}=g_{\vec{n}}\mathcal{R}^m$ and effective dispersion relation $E_{\left|\vec{n},q\right>}=-2J^{\vec{n}}\mathcal{F}_{1}\cos(q)+h \nu_r(n_x+x_y+1)$.   $q\in(-\pi,\pi]$ is the quasi-momentum and $J^{\vec{n}}$ is the nearest hopping.

The second term of Eq.(\ref{eq13}) can be taken as atoms oscillated between state $\left|\vec{n},q,g\right>$ and $\left|\vec{n},q+\Phi,e\right>$. Besides the Floquet sideband term $m\nu_s$, the detuning should be added an additional term $(E_{\left|\vec{n},q+\Phi\right>}-E_{\left|\vec{n},q\right>})/h$. Therefore, it is easy to derive the excited state probability of ground state atom  for $m$th Floquet sideband in $q$ state:
\begin{align}
	P_e^{\vec{n},m,q}(\delta,t)=\frac{g_{\vec{n},m}^2}{g_{\vec{n},m}^2+\delta_{\vec{n},m,q}^2}\sin^2(\sqrt{g_{\vec{n},m}^2+\delta_{\vec{n},m,q}^2}\pi t)\label{eq14}
\end{align}
where $\delta_{\vec{n},m,q}=\delta+m\nu_s+4J^{\vec{n}}\mathcal{F}_{1}\sin(q+\Phi/2)\sin(\Phi/2)/h$. One could see that for $\Phi\ne 0$, the resonance point that $\delta_{\vec{n},m,q}=0$ depends on $q$, thus atoms could be excited to certain quasi-momentum $q$ by adjusting the laser detuning. So spin orbit coupling phase $\Phi$ plays a crucial role in atom preparation.
\begin{figure}[h]
	\subfigure{\label{Rabi1}\includegraphics[width=0.49\linewidth]{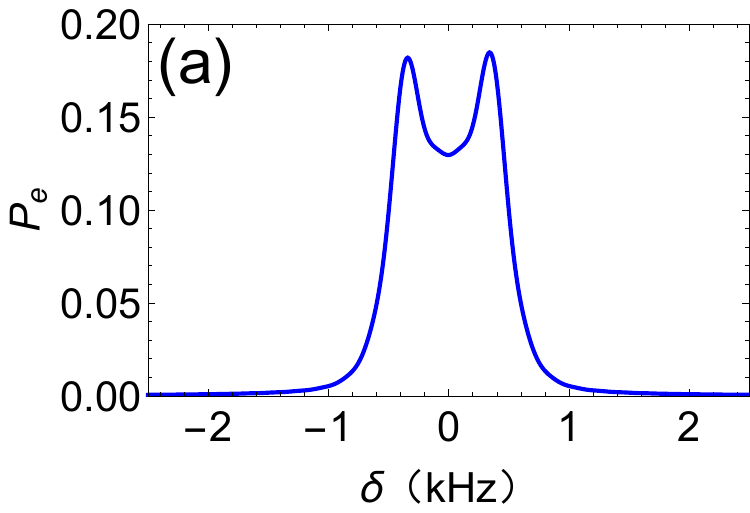}}
	\subfigure{\label{Rabi2}\includegraphics[width=0.49\linewidth]{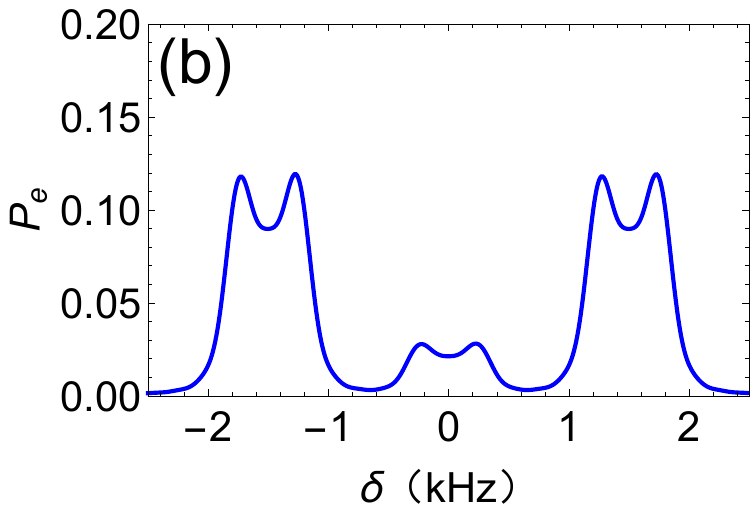}}
	\caption{Rabi spectrum with $J^{n_z}/h=80$Hz, $g_0=80$Hz, $C^{\vec{n}}/h=0.1(n_x+n_y+1)$Hz, $\nu_r=100$Hz, $t=5$ms, $N_x+N_y=2000$, lattice site number N=1000 and temperature $T=1\mu k$. (a) without driven. (b) periodic driving: $\nu_a=5000$ Hz and $\nu_s=1500$Hz.}
	\label{Rabi}.
\end{figure}

Considering the the Boltzmann distribution and using the resolved Floquet sideband approximation \cite{RSBA}, the Rabi spectrum of a large mount of atoms in OLC is:
\begin{eqnarray}
	P_e(\delta,t)=\sum_{q,m,\vec{n}}\frac{B(\vec{n},q)}{Z}P_e^{\vec{n},m,q}(\delta,t)\label{eq15}
\end{eqnarray}
in which $B(\vec{n},q)$ is the Boltzmann factor and $Z$ is partition function.

For cosine driving function $\Delta\nu(t)=\nu_a\cos(2\pi\nu_st)$, the Rabi spectrum is shown in Fig.\ref{Rabi2}, where $m=1,0,-1$ is considered and the effective coefficients are the first kind of Bessel functions $\mathcal{R}^m=\mathcal{J}_{m}(-\frac{\Phi\nu_a}{2\pi\nu_s})$, $\mathcal{F}_{1}=\mathcal{J}_n(\pi\frac{h\nu_a}{4E_r})$ with the lattice recoil energy $E_r=\hbar^2k_L^2/(2M)$. In our paper, we only focus on the lowest Bloch band ($n_z=0$) and ignore the inter-band transition for simplification.  

Eq.(\ref{eq14}) and Eq.(\ref{eq15}) manifest that the distribution of excited atoms relates to its quasi-momentum $q$, thus one can prepare atoms with precise momentum-dependent distributions by altering the detuning. After applying a laser with specific detuning $\delta_1$, distribution of atoms in momentum space could be written as 
\begin{align}
	\mathcal{P}_e(q,t_{p1})=\sum_{\vec{n}}\frac{B(\vec{n},q)}{Z}P_e^{\vec{n},m,q}(\delta_1,t_{p1}),\label{eq16}
\end{align}
where $m$ is the order of Floquet side band that one chose, $t_{p1}$ is the interaction time of the clock laser. 

\begin{figure}[htpb]
	\subfigure{\label{q-depend1}\includegraphics[width=0.49\linewidth]{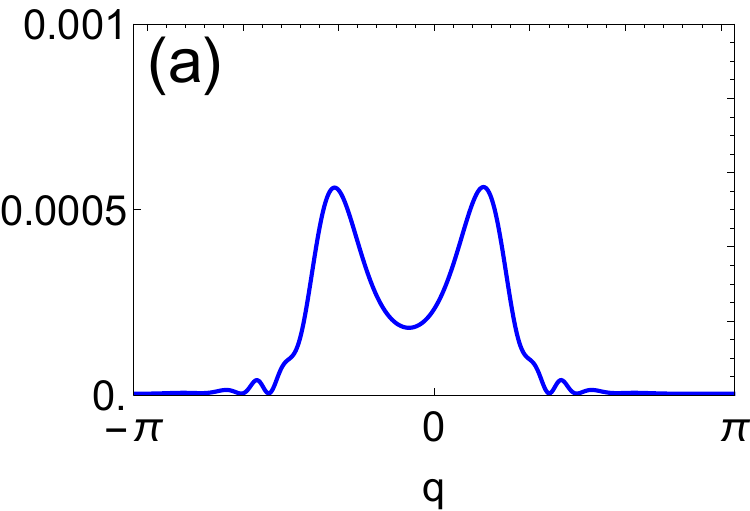}}
	\subfigure{\label{q-depend2}\includegraphics[width=0.49\linewidth]{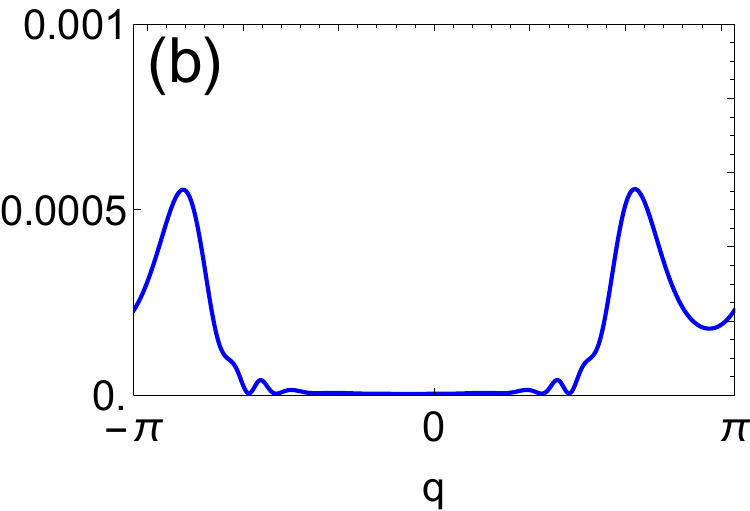}}
	\subfigure{\label{q-depend3}\includegraphics[width=0.49\linewidth]{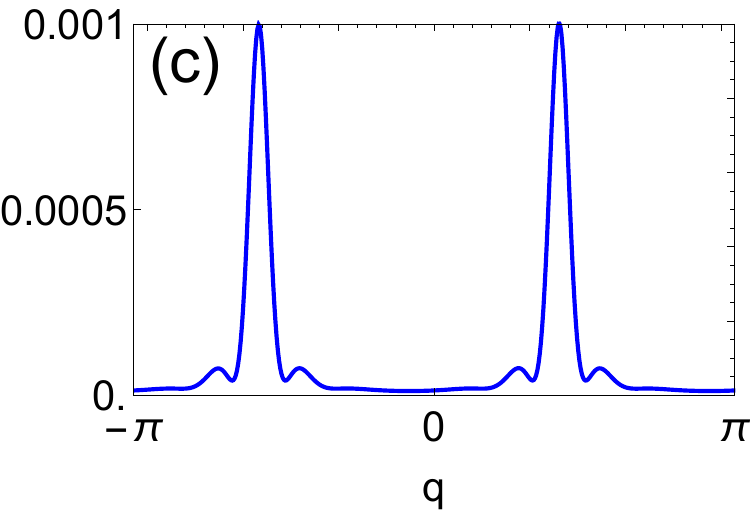}}
	\subfigure{\label{q-depend4}\includegraphics[width=0.49\linewidth]{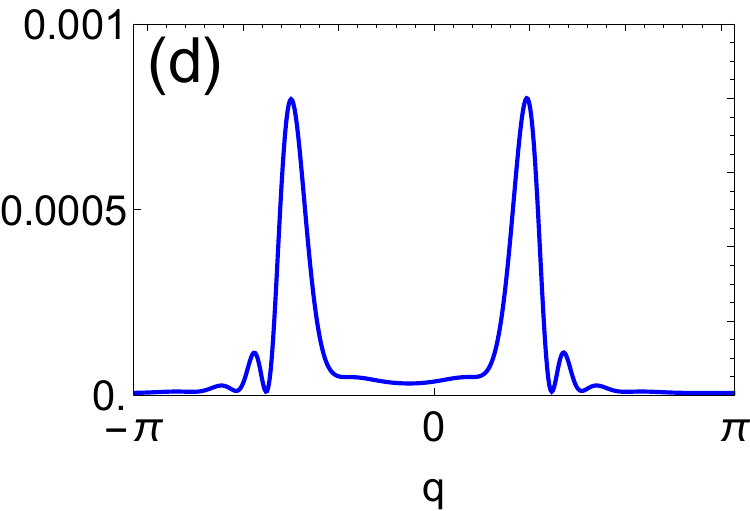}}
	\caption{Atoms preparation with momentum-dependent distribution (There is no division by $P_e(\delta_1,t_{p1})$ to normalize). (a) $\delta_1=\nu_s-4J^{n_z}\mathcal{F}_{1}\sin(\Phi/2)$. (b) $\delta_1=\nu_s+4J^{n_z}\mathcal{F}_{1}\sin(\Phi/2)$. (c) $\delta_1=\nu_s$. (d) $\delta_1=\nu_s-2J^{n_z}\mathcal{F}_{1}\sin(\Phi/2)$}
	\label{q-depend}.
\end{figure}

We chose $m=-1$th Floquet side band for example. Using a $\pi$ pulse clock laser, the atoms clustered in particular momentum is prepared by adjusting the detuning of clock laser, as shown in Fig. \ref{q-depend}. Here we chose the $\pi$ pulse to prepare as much atoms to $^{3}P_0$ state. After that, all the atoms remain in the ground states are kicked out by a cleaning laser.


\section{Rabi Spectroscopy of SBOs\label{sec4}} 
After the preparation of atoms to certain $q$, the clock laser is turned off. For stroboscopic detecting,  atoms could be seen as moving in momentum space in a constant velocity due to the effective force. The quasi-momentum changing after SBOs time $t$ is $\delta q(t)=2\pi \Delta d\nu_s t$, in which $d$ is the lattice constant. After SBOs, a second pulse clock laser is applied to bring atoms back to $^{1}S_0$ state.
The possibility of ground state for final momentum $q(t)=q+\delta q(t)$ is:
\begin{align}
	P_g^{\vec{n},m,q+q(t)}(\delta_2,t_{p2})=&\frac{g_{\vec{n},m}^2}{g_{\vec{n},m}^2+\delta_{\vec{n},m,q(t)}^2}\nonumber\\
	&\times\sin^2(\sqrt{g_{\vec{n},m}^2+\delta_{\vec{n},m,q(t)}^2}\pi t_{p2}),\label{eq18}
 \end{align}
where $\delta_{\vec{n},m,q(t)}=\delta_2+m\nu_s+(E_{\left|\vec{n},q(t)+\Phi\right>}-E_{\left|\vec{n},q(t)\right>})/h$ and $t_p$ is the probing time of the second clock laser. The final distribution of atoms in ground state $^{1}S_0$ also depends on the preparation, which is 

\begin{align}
	P_g(\delta_1,t_{p1};\delta_2,t_{p2})=&\frac{1}{P_e(\delta_1,t_{p1})}\sum_{\vec{n},q}\frac{B(\vec{n},q)}{Z}P_e^{\vec{n},m_1,q}(\delta_1,t_{p1}) \nonumber\\
		&\times P_g^{\vec{n},m_2,q+q(t)}(\delta_2,t_{p2}).\label{eq19}
\end{align}

\begin{figure}
	\centering
	\subfigure{\label{SBO1}\includegraphics[width=0.494\linewidth]{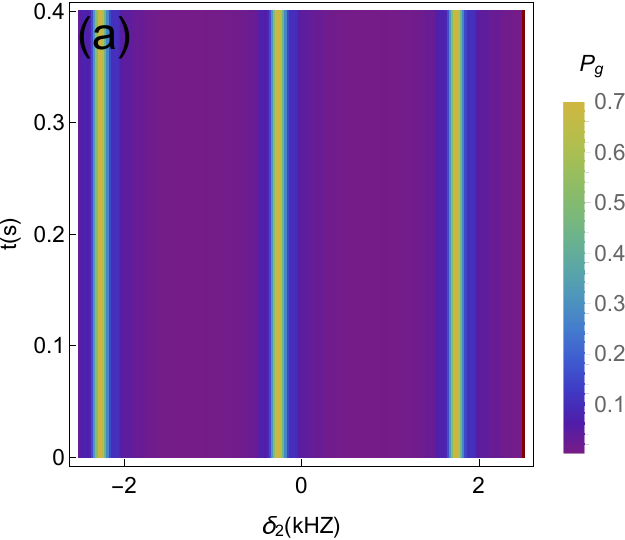}}
	\subfigure{\label{SBO2}\includegraphics[width=0.494\linewidth]{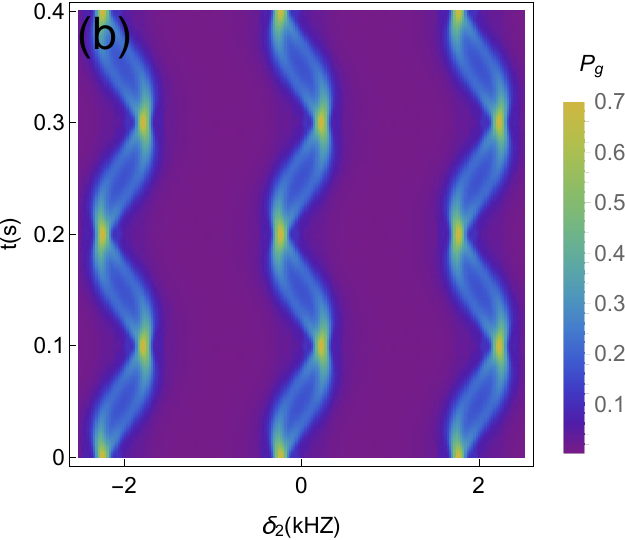}}
	\subfigure{\label{SBO3}\includegraphics[width=0.32\linewidth]{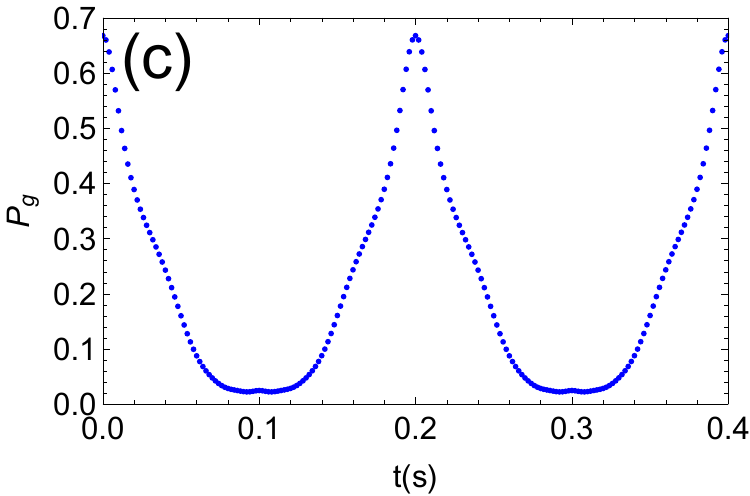}}
	\subfigure{\label{SBO4}\includegraphics[width=0.32\linewidth]{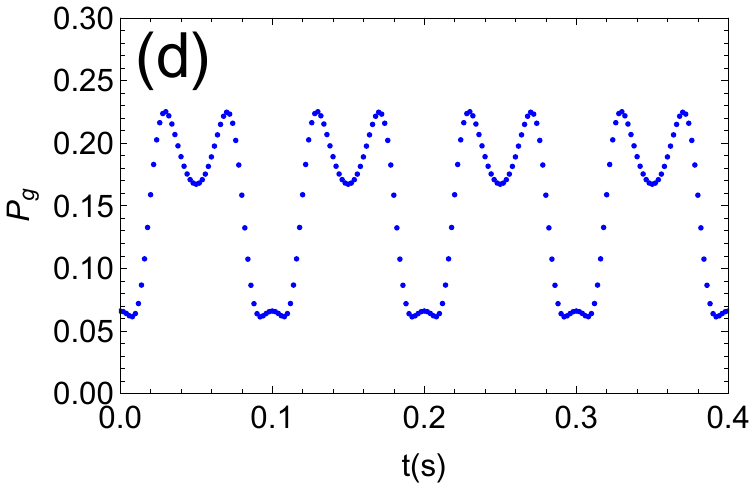}}
	\subfigure{\label{SBO5}\includegraphics[width=0.32\linewidth]{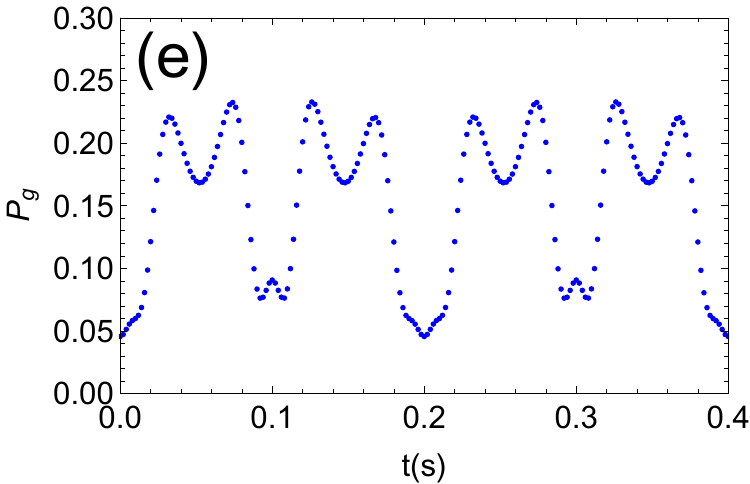}}
	\caption{(a)-(b): Rabi spectrum changing with waiting time $t$, where $J^{n_z}/h=120$Hz, $g_0=120$Hz, $\nu_s=2000$Hz, $n=1$, and other parameters are same as Fig.\ref{Rabi2}. It induces $g_{\vec{n},m}\approx65$Hz for $m=1,0,-1$ much bigger than $\Delta\nu_s=5$Hz, leading our approximation to be valid. The detuning $\delta$ of first clock laser is $\nu_s-4J^{n_z}\mathcal{F}_{1}\sin(\Phi/2)$ and the interacting time of two clock lasers is  $0.5/g_{\vec{n},-1}$. (a) resonant case. (b) off-resonant case with $\Delta=5/2000$. (c)-(e): The possibility of ground state $P_g$ changing with waiting time $t$ for different $\delta_2$ in off-resonant case. (c) $\delta_2=-2250$Hz. (d) $\delta_2=0$Hz. (e) $\delta_2=25$Hz.}
	\label{SBO}
\end{figure}

Scanning the detuning of the second laser $\delta_2$, one could get the Rabi spectrum. Here we presented a concrete example shown in Fig.\ref{SBO1} -\ref{SBO5} that atoms are prepared with $\delta_1=\nu_s-4J^{n_z}\mathcal{F}_{1}\sin(\Phi/2)$. Both $t_{p1}$ and $t_{p2}$ are chosen to be the corresponding $\pi$ pulse time. Fig.\ref{SBO1} represents the resonance case where $\Delta=0$, i.e., no effective force, there is no change in Rabi spectrum as atoms are static in quasi-momentum space. Fig.\ref{SBO2} represents the non resonance case where there is a small effective force, an obvious  oscillation of the Rabi spectrum is observed in each Floquet side band. To make it more clear, we also plot the ground state atom probability oscillating with time for a fixed detuning $\delta_2$, as shown in Figs.\ref{SBO3}-\ref{SBO5}. There is big difference among the oscillations for different detuning. Especially that the oscillating period reduce by half when $\delta_2=0$Hz (also for $m\nu_s$). The reason for this is that when $\delta_2=m\nu_s$, under the resolved Floquet side band approximation, $\delta_{\vec{n},m,q(t)}^2 \propto \sin^2(q(t))$, resulting in a half period.

OLC is one of the most accurate instrument in the world. Using its very high precision spectroscopy to observe the atom oscillation in momentum space may produce more accurate results than observing it in real space, which is limited by the resolution of CCD.  Periodically driving the optical lattice could reduce a large force to a small effective force, thus avoiding Landau-Zener transition between different bands. By utilizing spectroscopy of SBOs in OLC, one can precisely measure the force in a wide range. We will discuss this in next section.


\section{application : precise measurement of gravitational acceleration }\label{sec5}
As mentioned in previous Section, the constant force could be provided by the gravity of atoms. In a vertical OLC, the value of the force $(n+\Delta)\nu_s$ is an constant:
\begin{eqnarray}
	(n+\Delta)\nu_s = \frac{Mg\lambda_L}{2h} = \frac{g/\lambda_L}{4E_r/h}\label{eq20},
\end{eqnarray}
where $g$ is the gravitational acceleration. Its value is 875.3Hz in the $^{87}$Sr OLC ($g=9.8$$\textrm{m}/\textrm{s}^2$, $\lambda_L=813.43$$\textrm{nm}$, $E_r/h=3441$Hz). If we fix $n=1$, select the appropriate drive frequency $\nu_s$ and then measure the rabi spectrum of the SBOs like Figs.\ref{SBO3}-\ref{SBO5} to determining $\Delta\nu_s$, the gravitational acceleration $g$ can be accurately measured.

According to the error propagation expression, the uncertainty of gravitational acceleration can be written as:
\begin{eqnarray}
	\frac{\delta g}{g}=\frac{\delta(\Delta\nu_s)}{(n+\Delta)\nu_s}\label{eq21}
\end{eqnarray} 
Because of constant value of $(n+\Delta)\nu_s$, the smaller the error $\delta(\Delta\nu_s)$ is, the higher accuracy we can get.
The impact on the detection accuracy is mainly generated in three procedures
: state preparation, evolution and observation \cite{sensing}. Corresponding to our scheme, the Rabi frequency and the detuning of the two clock lasers as well as the SBOs time will affect the accuracy. To estimate the error of $\Delta\nu_s$, we calculate the Fisher information of the whole process:
\begin{eqnarray}
	F(\theta)=\frac{N_aP_e}{P_g(1-P_g)}\left( \frac{\partial P_g(\theta)}{\partial \theta}\right)^2\label{eq22},
\end{eqnarray}
where $N_a$ is the initial number of atoms. $P_e$ and $P_g$ can be obtained from Eq.(\ref{eq15}) and Eq.(\ref{eq19}), respectively. $\theta=\Delta\nu_s$ is the parameter we need to estimate. The inverse of the fisher information is an upper bound on the accuracy of the parameter estimates, which is known as Cramer-Rao theory \cite{C-R1,C-R2}: $\delta\theta\geq1/\sqrt{F(\theta)}$. Therefore, we need to find the conditions that make the maximum value of Eq.(\ref{eq22}) to determine the best measurement scheme. There is an intuitive result: fisher information becomes the larger with longer SBO time $t$ in the same condition, since $\partial P_g/\partial(\Delta\nu_s)$ will yield a factor of $t$. 
\begin{figure}
	\centering
	\subfigure{\includegraphics[width=0.494\linewidth]{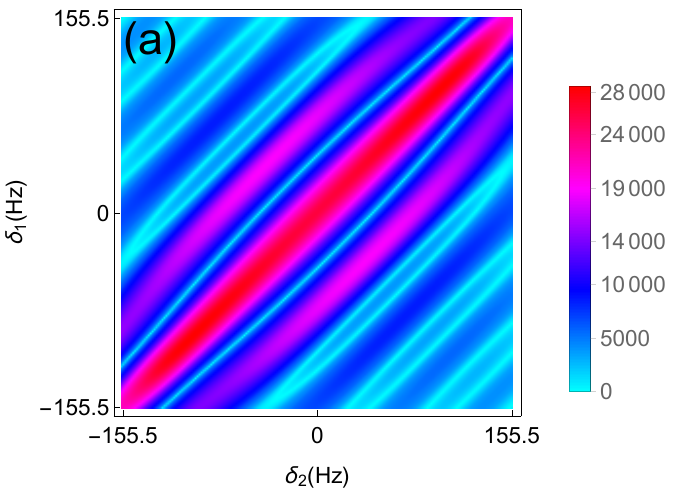}}
	\subfigure{\includegraphics[width=0.494\linewidth]{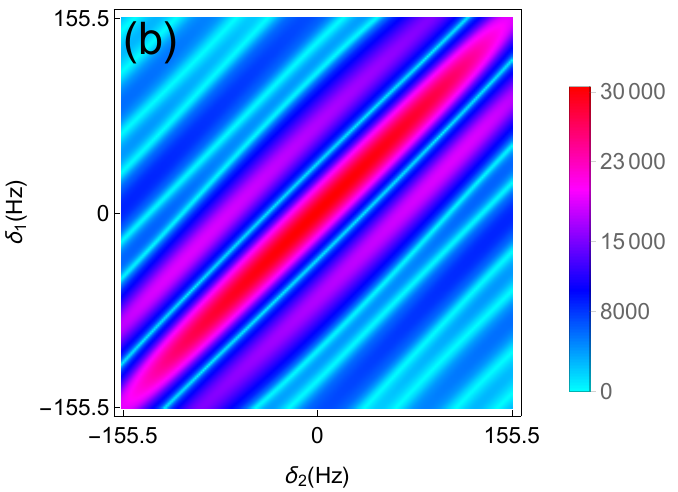}}
	\subfigure{\includegraphics[width=0.494\linewidth]{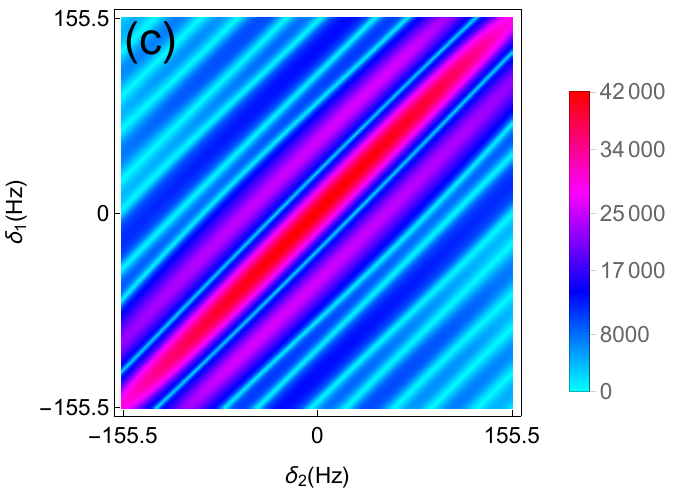}}
	\subfigure{\includegraphics[width=0.494\linewidth]{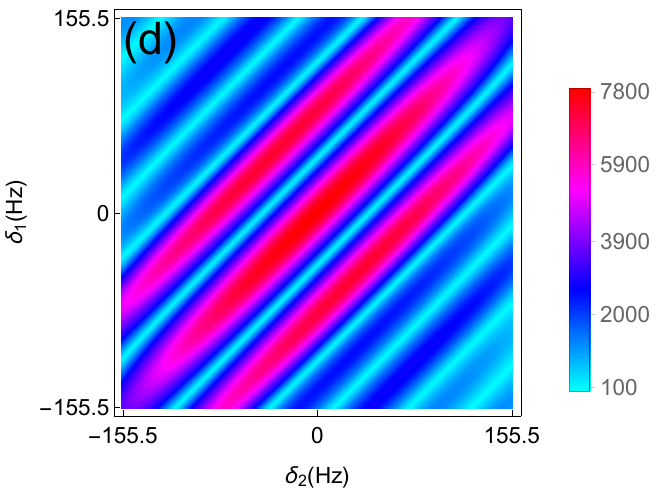}}
	\caption{Fisher information of different effective Rabi frequency and SBOs time with $J^{n_z}/h=80$Hz, $\mathcal{F}_{1}$=0.5, $\nu_s=874.3$Hz, $\Delta\nu_s=1$Hz, $C^{\vec{n}}/h=0.1(n_x+n_y+1)$Hz, $N_x+N_y=500$, $N_a=1$, lattice site number N=1000 and temperature $T=1\mu k$. (a) $g_1=g_2=30$Hz, $t=845T_s+19$s; (b) $g_1=40$Hz, $g_2=30$Hz, $t=850T_s+19$s; (c) $g_1=23$Hz, $g_2=33$Hz, $t=855T_s+19$s; (d) $g_1=52$Hz, $g_2=25$Hz, $t=860T_s+19$s. The interaction time of the first(second) clock laser is $0.5/g_1(g_2)$ and $4J^{n_z}\sin(\Phi/2)/h\approx155.5$Hz.}
	\label{fisher}
\end{figure}

\begin{figure}
	\includegraphics[width=0.99\linewidth]{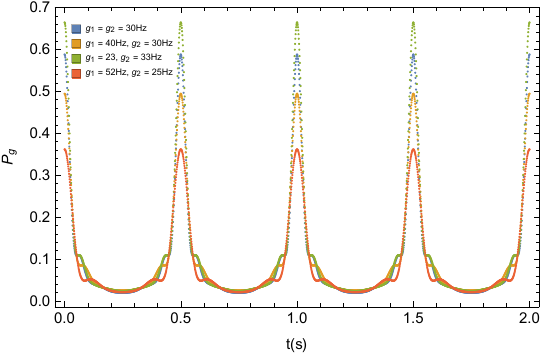}
	\caption{Rabi Spectroscopy of SBOs in the condition of $\delta_1=\delta_2=0$Hz. Other parameters are same as those in Fig.\ref{fisher}.}
	\label{SBOd=0}
\end{figure}

\begin{figure}
	\subfigure{\label{fig7a}\includegraphics[width=0.49\linewidth]{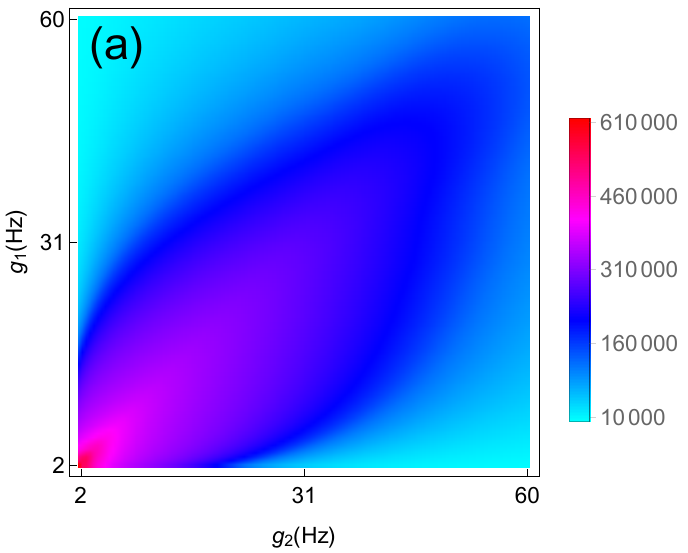}}
	\subfigure{\label{fig7b}\includegraphics[width=0.49\linewidth]{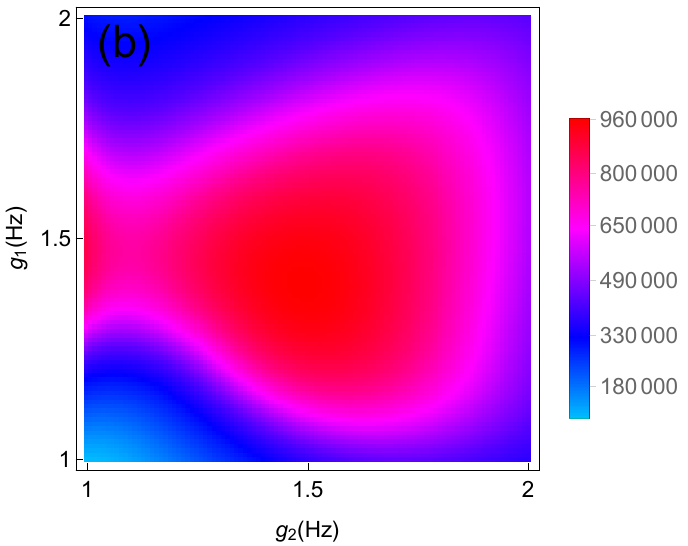}}
	\caption{Maximum Fisher information of $t=t_m+10$s ($t_m$ is the SBOs time with the Maximum Fisher information between $0.5/\Delta\nu_s$ and $1/\Delta\nu_s$) with $\nu_s=875.2$Hz, $\Delta\nu_s=0.1$Hz, $\delta_1=\delta_2=0$Hz and other parameters are same as those in Fig.\ref{fisher}.}
	\label{fisherg}
\end{figure}

To determine the optimal detuning and Rabi frequencies of two clock lasers, we calculated the Fisher information numerically according to Eq.(\ref{eq22}).  We first focus on the laser detuning. For simplicity, we restrict the detuning $\delta_1$ and $\delta_2$ to $0$th Floquet sideband because other cases are similar if we renormalize system and driving parameters. The effective rabi frequencies are also simplified to  $g_1$ and $g_2$ with $g_{1(2)}=g_{\vec{n},m}$ because of their tunability. As shown in Fig.\ref{fisher}, we calculate the Fisher information changing with $\delta_1$ and $\delta_2$ for several different $g_{1(2)}$. It turns out that the Fisher information of different $g_1$($g_2$) and $t$ is always symmetric with the point $\delta_1=\delta_2=0$Hz, whose Fisher information is always the biggest. Next we fix $\delta_1=\delta_2=0$ to search the optimal effective Rabi frequencies. The oscillation of this point with different $g_1$($g_2$) are shown in Fig.\ref{SBOd=0}, the line shape of which around $t=0.5k/\Delta\nu_s$ is similar to the Lorentzian function. Thus we could roughly estimate them by comparing their full width at half Maximum (FWHM). From Fig.\ref{SBOd=0} one could see that the smaller $g_1$($g_2$) are, the narrower FWHM of the spectrum is. This is further proved by numerically calculated the Maximum Fisher information with each $g_1$($g_2$) point, as shown in Fig.\ref{fig7a}, smaller effective Rabi frequencies $g_1$ and $g_2$ will result in a larger Fisher information. Here we give a brief explanation of why zero detuning and smaller effective Rabi frequencies results in larger Fisher information. Actually, $\delta_1=0$Hz and smaller $g_1$ correspond to the higher concentration of the momentum distribution exhibited in Fig. \ref{q-depend}. The smaller $g_{1(2)}$ is compared to $J^{\vec{n}}/h$, the more concentrated the momentum will be. Since the momentum is more concentrated, the diffusion of momentum caused by SBO will be very sharper and more sensitive to its changes. This is also why the second clock laser needs the same conditions. However, as $g_1$ becomes smaller, the number of prepared atoms $N_aP_e$ will also become smaller, leading to a decrease in Fisher information according to Eq.(\ref{eq22}). Thus, there is eventually an optimal $g_1$ due to the competition between momentum concentration and the number of prepared atoms, as shown in Fig.\ref{fig7b}. With our given parameters, the optimal $g_{1(2)}$ is around 1.5Hz, which satisfies our preconditions $g_{1(2)}\gg\Delta\nu_s$.

The experimental scheme achieved the best detection accuracy can be briefly summarized as: i) the longer SBOs time $t$; ii) the detuning of two clock lasers $\delta_1=\delta_2=0$Hz; iii) Numerically, there exists an optimal effective Rabi frequency $g_{1(2)}$ if the optimal numerical result satisfies the condition $g_{1(2)}\gg\Delta\nu_s$. Otherwise, smaller $g_{1(2)}$ is better under the condition $g_{1(2)}\gg\Delta\nu_s$. Considering that the excited state lifetime of $^{87}$Sr can reach 128s, we calculate the theoretical accuracy that can be achieved by using OLC: $\delta g/g\approx6\times 10^{-10}$ with $N_a=10^5$, $t=t_m+110$s, $g_1=g_2=1.5$ Hz and other same parameters in Fig.\ref{fisherg}. There are also limits of driving frequency. To let effective Hamiltonian Eq.(\ref{eq9}) be held, the characteristic energy $h\nu_s$ should be much smaller than band gap as well as much larger than effective hopping strength, this usually gives the driving frequency parameter window to be several hundreds to several Kilo Hertz.
\section{conclusion and outlook}
In summary, we have derived the effective model of OLC with atoms been added both static and time-periodical forces. Under appropriate parameter conditions, atoms could be seen as moving in the momentum space with a constant velocity by a renormalized static force, which is the so called SBOs. Thanks to the spin orbit coupling phase $\Phi$, the atoms could be prepared to clustered around certain momentum. By controlling driving frequency, a large static force could be reduced to a small force that Landau Zener transition to other bands could be avoided. The Rabi spectroscopy of SBOs for OLC under experimental achievable condition is given. Using precise spectroscopy of OLC, force with a large range could be measured. The best measurement scheme is also discussed by calculating Fisher information numerically.

Lots of work could be done based on this paper in future. One work would be replacing the periodically driving force by a quasi-periodical one, together with the static force, a directed SBOs could be observed\cite{Yuce_2013}. Another direction would be utilizing the precision of OLC to quantize the Landau Zener transition between Floquet dressed energy bands, which will provide accurate data for heating problems of Floquet engineering\cite{PhysRevX.10.021044}.  
\section{ACKNOWLEDGMENTS}
This work is supported by the National Science Foundation
of China under Grants No. 12274045, No. 11874094, No.12147102.  T. Wang acknowledges funding supported by the Program of State Key Laboratory of Quantum Optics and Quantum Optics Devices(No:KF202211) and Fundamental Research Funds for the Central Universities Grant No. 2023CDJXY-048. Y. zhang acknowledges National Science Foundation
of China under Grants No.52071121. 
\bibliographystyle{apsrev4-2}
\bibliography{ref}

\begin{thebibliography}{24}%
\makeatletter
\providecommand \@ifxundefined [1]{%
 \@ifx{#1\undefined}
}%
\providecommand \@ifnum [1]{%
 \ifnum #1\expandafter \@firstoftwo
 \else \expandafter \@secondoftwo
 \fi
}%
\providecommand \@ifx [1]{%
 \ifx #1\expandafter \@firstoftwo
 \else \expandafter \@secondoftwo
 \fi
}%
\providecommand \natexlab [1]{#1}%
\providecommand \enquote  [1]{``#1''}%
\providecommand \bibnamefont  [1]{#1}%
\providecommand \bibfnamefont [1]{#1}%
\providecommand \citenamefont [1]{#1}%
\providecommand \href@noop [0]{\@secondoftwo}%
\providecommand \href [0]{\begingroup \@sanitize@url \@href}%
\providecommand \@href[1]{\@@startlink{#1}\@@href}%
\providecommand \@@href[1]{\endgroup#1\@@endlink}%
\providecommand \@sanitize@url [0]{\catcode `\\12\catcode `\$12\catcode
  `\&12\catcode `\#12\catcode `\^12\catcode `\_12\catcode `\%12\relax}%
\providecommand \@@startlink[1]{}%
\providecommand \@@endlink[0]{}%
\providecommand \url  [0]{\begingroup\@sanitize@url \@url }%
\providecommand \@url [1]{\endgroup\@href {#1}{\urlprefix }}%
\providecommand \urlprefix  [0]{URL }%
\providecommand \Eprint [0]{\href }%
\providecommand \doibase [0]{https://doi.org/}%
\providecommand \selectlanguage [0]{\@gobble}%
\providecommand \bibinfo  [0]{\@secondoftwo}%
\providecommand \bibfield  [0]{\@secondoftwo}%
\providecommand \translation [1]{[#1]}%
\providecommand \BibitemOpen [0]{}%
\providecommand \bibitemStop [0]{}%
\providecommand \bibitemNoStop [0]{.\EOS\space}%
\providecommand \EOS [0]{\spacefactor3000\relax}%
\providecommand \BibitemShut  [1]{\csname bibitem#1\endcsname}%
\let\auto@bib@innerbib\@empty
\bibitem [{\citenamefont {Bloch}(1929)}]{BO1}%
  \BibitemOpen
  \bibfield  {author} {\bibinfo {author} {\bibfnamefont {F.}~\bibnamefont
  {Bloch}},\ }\href {https://doi.org/10.1007/BF01339455} {\bibfield  {journal}
  {\bibinfo  {journal} {Z. Physik}\ }\textbf {\bibinfo {volume} {52}},\
  \bibinfo {pages} {555} (\bibinfo {year} {1929})}\BibitemShut {NoStop}%
\bibitem [{\citenamefont {Zener}(1934)}]{BO2}%
  \BibitemOpen
  \bibfield  {author} {\bibinfo {author} {\bibfnamefont {C.}~\bibnamefont
  {Zener}},\ }\href {https://doi.org/10.1098/rspa.1934.0116} {\bibfield
  {journal} {\bibinfo  {journal} {Proc. R. Soc. Lond. A}\ }\textbf {\bibinfo
  {volume} {145}},\ \bibinfo {pages} {523} (\bibinfo {year}
  {1934})}\BibitemShut {NoStop}%
\bibitem [{\citenamefont {Feldmann}\ \emph {et~al.}(1992)\citenamefont
  {Feldmann}, \citenamefont {Leo}, \citenamefont {Shah}, \citenamefont
  {Miller}, \citenamefont {Cunningham}, \citenamefont {Meier}, \citenamefont
  {von Plessen}, \citenamefont {Schulze}, \citenamefont {Thomas},\ and\
  \citenamefont {Schmitt-Rink}}]{superlattice1}%
  \BibitemOpen
  \bibfield  {author} {\bibinfo {author} {\bibfnamefont {J.}~\bibnamefont
  {Feldmann}}, \bibinfo {author} {\bibfnamefont {K.}~\bibnamefont {Leo}},
  \bibinfo {author} {\bibfnamefont {J.}~\bibnamefont {Shah}}, \bibinfo {author}
  {\bibfnamefont {D.~A.~B.}\ \bibnamefont {Miller}}, \bibinfo {author}
  {\bibfnamefont {J.~E.}\ \bibnamefont {Cunningham}}, \bibinfo {author}
  {\bibfnamefont {T.}~\bibnamefont {Meier}}, \bibinfo {author} {\bibfnamefont
  {G.}~\bibnamefont {von Plessen}}, \bibinfo {author} {\bibfnamefont
  {A.}~\bibnamefont {Schulze}}, \bibinfo {author} {\bibfnamefont
  {P.}~\bibnamefont {Thomas}},\ and\ \bibinfo {author} {\bibfnamefont
  {S.}~\bibnamefont {Schmitt-Rink}},\ }\href
  {https://doi.org/10.1103/PhysRevB.46.7252} {\bibfield  {journal} {\bibinfo
  {journal} {Phys. Rev. B}\ }\textbf {\bibinfo {volume} {46}},\ \bibinfo
  {pages} {7252} (\bibinfo {year} {1992})}\BibitemShut {NoStop}%
\bibitem [{\citenamefont {Leo}\ \emph {et~al.}(1992)\citenamefont {Leo},
  \citenamefont {Bolivar}, \citenamefont {Brüggemann}, \citenamefont
  {Schwedler},\ and\ \citenamefont {Köhler}}]{superlattice2}%
  \BibitemOpen
  \bibfield  {author} {\bibinfo {author} {\bibfnamefont {K.}~\bibnamefont
  {Leo}}, \bibinfo {author} {\bibfnamefont {P.~H.}\ \bibnamefont {Bolivar}},
  \bibinfo {author} {\bibfnamefont {F.}~\bibnamefont {Brüggemann}}, \bibinfo
  {author} {\bibfnamefont {R.}~\bibnamefont {Schwedler}},\ and\ \bibinfo
  {author} {\bibfnamefont {K.}~\bibnamefont {Köhler}},\ }\href
  {https://doi.org/https://doi.org/10.1016/0038-1098(92)90798-E} {\bibfield
  {journal} {\bibinfo  {journal} {Solid State Communications}\ }\textbf
  {\bibinfo {volume} {84}},\ \bibinfo {pages} {943} (\bibinfo {year}
  {1992})}\BibitemShut {NoStop}%
\bibitem [{\citenamefont {Ben~Dahan}\ \emph {et~al.}(1996)\citenamefont
  {Ben~Dahan}, \citenamefont {Peik}, \citenamefont {Reichel}, \citenamefont
  {Castin},\ and\ \citenamefont {Salomon}}]{OL}%
  \BibitemOpen
  \bibfield  {author} {\bibinfo {author} {\bibfnamefont {M.}~\bibnamefont
  {Ben~Dahan}}, \bibinfo {author} {\bibfnamefont {E.}~\bibnamefont {Peik}},
  \bibinfo {author} {\bibfnamefont {J.}~\bibnamefont {Reichel}}, \bibinfo
  {author} {\bibfnamefont {Y.}~\bibnamefont {Castin}},\ and\ \bibinfo {author}
  {\bibfnamefont {C.}~\bibnamefont {Salomon}},\ }\href
  {https://doi.org/10.1103/PhysRevLett.76.4508} {\bibfield  {journal} {\bibinfo
   {journal} {Phys. Rev. Lett.}\ }\textbf {\bibinfo {volume} {76}},\ \bibinfo
  {pages} {4508} (\bibinfo {year} {1996})}\BibitemShut {NoStop}%
\bibitem [{\citenamefont {Kolkowitz}\ \emph {et~al.}(2017)\citenamefont
  {Kolkowitz}, \citenamefont {Bromley}, \citenamefont {Bothwell}, \citenamefont
  {Wall}, \citenamefont {Marti}, \citenamefont {Koller}, \citenamefont {Zhang},
  \citenamefont {Rey},\ and\ \citenamefont {Ye}}]{OLC}%
  \BibitemOpen
  \bibfield  {author} {\bibinfo {author} {\bibfnamefont {S.}~\bibnamefont
  {Kolkowitz}}, \bibinfo {author} {\bibfnamefont {S.~L.}\ \bibnamefont
  {Bromley}}, \bibinfo {author} {\bibfnamefont {T.}~\bibnamefont {Bothwell}},
  \bibinfo {author} {\bibfnamefont {M.~L.}\ \bibnamefont {Wall}}, \bibinfo
  {author} {\bibfnamefont {G.~E.}\ \bibnamefont {Marti}}, \bibinfo {author}
  {\bibfnamefont {A.~P.}\ \bibnamefont {Koller}}, \bibinfo {author}
  {\bibfnamefont {X.}~\bibnamefont {Zhang}}, \bibinfo {author} {\bibfnamefont
  {A.~M.}\ \bibnamefont {Rey}},\ and\ \bibinfo {author} {\bibfnamefont
  {J.}~\bibnamefont {Ye}},\ }\href {https://doi.org/10.1038/nature20811}
  {\bibfield  {journal} {\bibinfo  {journal} {Nature}\ }\textbf {\bibinfo
  {volume} {542}},\ \bibinfo {pages} {66–70} (\bibinfo {year}
  {2017})}\BibitemShut {NoStop}%
\bibitem [{\citenamefont {Guo}\ \emph {et~al.}(2021)\citenamefont {Guo},
  \citenamefont {Ge}, \citenamefont {Li}, \citenamefont {Wang}, \citenamefont
  {Zhang}, \citenamefont {Song}, \citenamefont {Xiang}, \citenamefont {Song},
  \citenamefont {Jin}, \citenamefont {Lu}, \citenamefont {Xu}, \citenamefont
  {Zheng},\ and\ \citenamefont {Fan}}]{SQB}%
  \BibitemOpen
  \bibfield  {author} {\bibinfo {author} {\bibfnamefont {X.-Y.}\ \bibnamefont
  {Guo}}, \bibinfo {author} {\bibfnamefont {Z.-Y.}\ \bibnamefont {Ge}},
  \bibinfo {author} {\bibfnamefont {H.}~\bibnamefont {Li}}, \bibinfo {author}
  {\bibfnamefont {Z.}~\bibnamefont {Wang}}, \bibinfo {author} {\bibfnamefont
  {Y.-R.}\ \bibnamefont {Zhang}}, \bibinfo {author} {\bibfnamefont
  {P.}~\bibnamefont {Song}}, \bibinfo {author} {\bibfnamefont {Z.}~\bibnamefont
  {Xiang}}, \bibinfo {author} {\bibfnamefont {X.}~\bibnamefont {Song}},
  \bibinfo {author} {\bibfnamefont {Y.}~\bibnamefont {Jin}}, \bibinfo {author}
  {\bibfnamefont {L.}~\bibnamefont {Lu}}, \bibinfo {author} {\bibfnamefont
  {K.}~\bibnamefont {Xu}}, \bibinfo {author} {\bibfnamefont {D.}~\bibnamefont
  {Zheng}},\ and\ \bibinfo {author} {\bibfnamefont {H.}~\bibnamefont {Fan}},\
  }\href {https://doi.org/10.1038/s41534-021-00385-3} {\bibfield  {journal}
  {\bibinfo  {journal} {npj Quantum Information}\ }\textbf {\bibinfo {volume}
  {7}},\ \bibinfo {pages} {51} (\bibinfo {year} {2021})}\BibitemShut {NoStop}%
\bibitem [{\citenamefont {Fujiwara}\ \emph {et~al.}(2019)\citenamefont
  {Fujiwara}, \citenamefont {Singh}, \citenamefont {Geiger}, \citenamefont
  {Senaratne}, \citenamefont {Rajagopal}, \citenamefont {Lipatov},\ and\
  \citenamefont {Weld}}]{F-B1}%
  \BibitemOpen
  \bibfield  {author} {\bibinfo {author} {\bibfnamefont {C.~J.}\ \bibnamefont
  {Fujiwara}}, \bibinfo {author} {\bibfnamefont {K.}~\bibnamefont {Singh}},
  \bibinfo {author} {\bibfnamefont {Z.~A.}\ \bibnamefont {Geiger}}, \bibinfo
  {author} {\bibfnamefont {R.}~\bibnamefont {Senaratne}}, \bibinfo {author}
  {\bibfnamefont {S.~V.}\ \bibnamefont {Rajagopal}}, \bibinfo {author}
  {\bibfnamefont {M.}~\bibnamefont {Lipatov}},\ and\ \bibinfo {author}
  {\bibfnamefont {D.~M.}\ \bibnamefont {Weld}},\ }\href
  {https://doi.org/10.1103/PhysRevLett.122.010402} {\bibfield  {journal}
  {\bibinfo  {journal} {Phys. Rev. Lett.}\ }\textbf {\bibinfo {volume} {122}},\
  \bibinfo {pages} {010402} (\bibinfo {year} {2019})}\BibitemShut {NoStop}%
\bibitem [{\citenamefont {Sandholzer}\ \emph {et~al.}(2022)\citenamefont
  {Sandholzer}, \citenamefont {Walter}, \citenamefont {Minguzzi}, \citenamefont
  {Zhu}, \citenamefont {Viebahn},\ and\ \citenamefont {Esslinger}}]{F-B2}%
  \BibitemOpen
  \bibfield  {author} {\bibinfo {author} {\bibfnamefont {K.}~\bibnamefont
  {Sandholzer}}, \bibinfo {author} {\bibfnamefont {A.-S.}\ \bibnamefont
  {Walter}}, \bibinfo {author} {\bibfnamefont {J.}~\bibnamefont {Minguzzi}},
  \bibinfo {author} {\bibfnamefont {Z.}~\bibnamefont {Zhu}}, \bibinfo {author}
  {\bibfnamefont {K.}~\bibnamefont {Viebahn}},\ and\ \bibinfo {author}
  {\bibfnamefont {T.}~\bibnamefont {Esslinger}},\ }\href
  {https://doi.org/10.1103/PhysRevResearch.4.013056} {\bibfield  {journal}
  {\bibinfo  {journal} {Phys. Rev. Res.}\ }\textbf {\bibinfo {volume} {4}},\
  \bibinfo {pages} {013056} (\bibinfo {year} {2022})}\BibitemShut {NoStop}%
\bibitem [{\citenamefont {Ferrari}\ \emph {et~al.}(2006)\citenamefont
  {Ferrari}, \citenamefont {Poli}, \citenamefont {Sorrentino},\ and\
  \citenamefont {Tino}}]{Gravity1}%
  \BibitemOpen
  \bibfield  {author} {\bibinfo {author} {\bibfnamefont {G.}~\bibnamefont
  {Ferrari}}, \bibinfo {author} {\bibfnamefont {N.}~\bibnamefont {Poli}},
  \bibinfo {author} {\bibfnamefont {F.}~\bibnamefont {Sorrentino}},\ and\
  \bibinfo {author} {\bibfnamefont {G.~M.}\ \bibnamefont {Tino}},\ }\href
  {https://doi.org/10.1103/PhysRevLett.97.060402} {\bibfield  {journal}
  {\bibinfo  {journal} {Phys. Rev. Lett.}\ }\textbf {\bibinfo {volume} {97}},\
  \bibinfo {pages} {060402} (\bibinfo {year} {2006})}\BibitemShut {NoStop}%
\bibitem [{\citenamefont {Poli}\ \emph {et~al.}(2011)\citenamefont {Poli},
  \citenamefont {Wang}, \citenamefont {Tarallo}, \citenamefont {Alberti},
  \citenamefont {Prevedelli},\ and\ \citenamefont {Tino}}]{Gravity2}%
  \BibitemOpen
  \bibfield  {author} {\bibinfo {author} {\bibfnamefont {N.}~\bibnamefont
  {Poli}}, \bibinfo {author} {\bibfnamefont {F.-Y.}\ \bibnamefont {Wang}},
  \bibinfo {author} {\bibfnamefont {M.~G.}\ \bibnamefont {Tarallo}}, \bibinfo
  {author} {\bibfnamefont {A.}~\bibnamefont {Alberti}}, \bibinfo {author}
  {\bibfnamefont {M.}~\bibnamefont {Prevedelli}},\ and\ \bibinfo {author}
  {\bibfnamefont {G.~M.}\ \bibnamefont {Tino}},\ }\href
  {https://doi.org/10.1103/PhysRevLett.106.038501} {\bibfield  {journal}
  {\bibinfo  {journal} {Phys. Rev. Lett.}\ }\textbf {\bibinfo {volume} {106}},\
  \bibinfo {pages} {038501} (\bibinfo {year} {2011})}\BibitemShut {NoStop}%
\bibitem [{\citenamefont {Haller}\ \emph {et~al.}(2010)\citenamefont {Haller},
  \citenamefont {Hart}, \citenamefont {Mark}, \citenamefont {Danzl},
  \citenamefont {Reichs\"ollner},\ and\ \citenamefont {N\"agerl}}]{SB0s-ep}%
  \BibitemOpen
  \bibfield  {author} {\bibinfo {author} {\bibfnamefont {E.}~\bibnamefont
  {Haller}}, \bibinfo {author} {\bibfnamefont {R.}~\bibnamefont {Hart}},
  \bibinfo {author} {\bibfnamefont {M.~J.}\ \bibnamefont {Mark}}, \bibinfo
  {author} {\bibfnamefont {J.~G.}\ \bibnamefont {Danzl}}, \bibinfo {author}
  {\bibfnamefont {L.}~\bibnamefont {Reichs\"ollner}},\ and\ \bibinfo {author}
  {\bibfnamefont {H.-C.}\ \bibnamefont {N\"agerl}},\ }\href
  {https://doi.org/10.1103/PhysRevLett.104.200403} {\bibfield  {journal}
  {\bibinfo  {journal} {Phys. Rev. Lett.}\ }\textbf {\bibinfo {volume} {104}},\
  \bibinfo {pages} {200403} (\bibinfo {year} {2010})}\BibitemShut {NoStop}%
\bibitem [{\citenamefont {Kudo}\ and\ \citenamefont
  {Monteiro}(2011)}]{SBOs-th1}%
  \BibitemOpen
  \bibfield  {author} {\bibinfo {author} {\bibfnamefont {K.}~\bibnamefont
  {Kudo}}\ and\ \bibinfo {author} {\bibfnamefont {T.~S.}\ \bibnamefont
  {Monteiro}},\ }\href {https://doi.org/10.1103/PhysRevA.83.053627} {\bibfield
  {journal} {\bibinfo  {journal} {Phys. Rev. A}\ }\textbf {\bibinfo {volume}
  {83}},\ \bibinfo {pages} {053627} (\bibinfo {year} {2011})}\BibitemShut
  {NoStop}%
\bibitem [{\citenamefont {Arlinghaus}\ and\ \citenamefont
  {Holthaus}(2011)}]{SBOs-th2}%
  \BibitemOpen
  \bibfield  {author} {\bibinfo {author} {\bibfnamefont {S.}~\bibnamefont
  {Arlinghaus}}\ and\ \bibinfo {author} {\bibfnamefont {M.}~\bibnamefont
  {Holthaus}},\ }\href {https://doi.org/10.1103/PhysRevB.84.054301} {\bibfield
  {journal} {\bibinfo  {journal} {Phys. Rev. B}\ }\textbf {\bibinfo {volume}
  {84}},\ \bibinfo {pages} {054301} (\bibinfo {year} {2011})}\BibitemShut
  {NoStop}%
\bibitem [{\citenamefont {Yin}\ \emph {et~al.}(2022)\citenamefont {Yin},
  \citenamefont {Lu}, \citenamefont {Li}, \citenamefont {Xia}, \citenamefont
  {Wang}, \citenamefont {Zhang},\ and\ \citenamefont {Chang}}]{FE-OLC}%
  \BibitemOpen
  \bibfield  {author} {\bibinfo {author} {\bibfnamefont {M.-J.}\ \bibnamefont
  {Yin}}, \bibinfo {author} {\bibfnamefont {X.-T.}\ \bibnamefont {Lu}},
  \bibinfo {author} {\bibfnamefont {T.}~\bibnamefont {Li}}, \bibinfo {author}
  {\bibfnamefont {J.-J.}\ \bibnamefont {Xia}}, \bibinfo {author} {\bibfnamefont
  {T.}~\bibnamefont {Wang}}, \bibinfo {author} {\bibfnamefont {X.-F.}\
  \bibnamefont {Zhang}},\ and\ \bibinfo {author} {\bibfnamefont
  {H.}~\bibnamefont {Chang}},\ }\href
  {https://doi.org/10.1103/PhysRevLett.128.073603} {\bibfield  {journal}
  {\bibinfo  {journal} {Phys. Rev. Lett.}\ }\textbf {\bibinfo {volume} {128}},\
  \bibinfo {pages} {073603} (\bibinfo {year} {2022})}\BibitemShut {NoStop}%
\bibitem [{\citenamefont {{Boulder Atomic Clock Optical Network (BACON)
  Collaboration*}}(2021)}]{frequency}%
  \BibitemOpen
  \bibfield  {author} {\bibinfo {author} {\bibnamefont {{Boulder Atomic Clock
  Optical Network (BACON) Collaboration*}}},\ }\href
  {https://doi.org/10.1038/s41586-021-03253-4} {\bibfield  {journal} {\bibinfo
  {journal} {Nature}\ }\textbf {\bibinfo {volume} {591}},\ \bibinfo {pages}
  {564} (\bibinfo {year} {2021})}\BibitemShut {NoStop}%
\bibitem [{\citenamefont {Eckardt}\ and\ \citenamefont
  {Holthaus}(2007)}]{Bessel1}%
  \BibitemOpen
  \bibfield  {author} {\bibinfo {author} {\bibfnamefont {A.}~\bibnamefont
  {Eckardt}}\ and\ \bibinfo {author} {\bibfnamefont {M.}~\bibnamefont
  {Holthaus}},\ }\href {https://doi.org/10.1209/0295-5075/80/50004} {\bibfield
  {journal} {\bibinfo  {journal} {Europhysics Letters ({EPL})}\ }\textbf
  {\bibinfo {volume} {80}},\ \bibinfo {pages} {50004} (\bibinfo {year}
  {2007})}\BibitemShut {NoStop}%
\bibitem [{\citenamefont {Eckardt}\ \emph {et~al.}(2005)\citenamefont
  {Eckardt}, \citenamefont {Weiss},\ and\ \citenamefont {Holthaus}}]{Bessel2}%
  \BibitemOpen
  \bibfield  {author} {\bibinfo {author} {\bibfnamefont {A.}~\bibnamefont
  {Eckardt}}, \bibinfo {author} {\bibfnamefont {C.}~\bibnamefont {Weiss}},\
  and\ \bibinfo {author} {\bibfnamefont {M.}~\bibnamefont {Holthaus}},\ }\href
  {https://doi.org/10.1103/PhysRevLett.95.260404} {\bibfield  {journal}
  {\bibinfo  {journal} {Phys. Rev. Lett.}\ }\textbf {\bibinfo {volume} {95}},\
  \bibinfo {pages} {260404} (\bibinfo {year} {2005})}\BibitemShut {NoStop}%
\bibitem [{\citenamefont {Yin}\ \emph {et~al.}(2021)\citenamefont {Yin},
  \citenamefont {Wang}, \citenamefont {Lu}, \citenamefont {Li}, \citenamefont
  {Wang}, \citenamefont {Zhang}, \citenamefont {Li}, \citenamefont {Smerzi},\
  and\ \citenamefont {Chang}}]{RSBA}%
  \BibitemOpen
  \bibfield  {author} {\bibinfo {author} {\bibfnamefont {M.-J.}\ \bibnamefont
  {Yin}}, \bibinfo {author} {\bibfnamefont {T.}~\bibnamefont {Wang}}, \bibinfo
  {author} {\bibfnamefont {X.-T.}\ \bibnamefont {Lu}}, \bibinfo {author}
  {\bibfnamefont {T.}~\bibnamefont {Li}}, \bibinfo {author} {\bibfnamefont
  {Y.-B.}\ \bibnamefont {Wang}}, \bibinfo {author} {\bibfnamefont {X.-F.}\
  \bibnamefont {Zhang}}, \bibinfo {author} {\bibfnamefont {W.-D.}\ \bibnamefont
  {Li}}, \bibinfo {author} {\bibfnamefont {A.}~\bibnamefont {Smerzi}},\ and\
  \bibinfo {author} {\bibfnamefont {H.}~\bibnamefont {Chang}},\ }\href
  {https://doi.org/10.1088/0256-307X/38/7/073201} {\bibfield  {journal}
  {\bibinfo  {journal} {Chinese Physics Letters}\ }\textbf {\bibinfo {volume}
  {38}},\ \bibinfo {eid} {073201} (\bibinfo {year} {2021})}\BibitemShut
  {NoStop}%
\bibitem [{\citenamefont {Degen}\ \emph {et~al.}(2017)\citenamefont {Degen},
  \citenamefont {Reinhard},\ and\ \citenamefont {Cappellaro}}]{sensing}%
  \BibitemOpen
  \bibfield  {author} {\bibinfo {author} {\bibfnamefont {C.~L.}\ \bibnamefont
  {Degen}}, \bibinfo {author} {\bibfnamefont {F.}~\bibnamefont {Reinhard}},\
  and\ \bibinfo {author} {\bibfnamefont {P.}~\bibnamefont {Cappellaro}},\
  }\href {https://doi.org/10.1103/RevModPhys.89.035002} {\bibfield  {journal}
  {\bibinfo  {journal} {Rev. Mod. Phys.}\ }\textbf {\bibinfo {volume} {89}},\
  \bibinfo {pages} {035002} (\bibinfo {year} {2017})}\BibitemShut {NoStop}%
\bibitem [{\citenamefont {Pezz\`e}\ \emph {et~al.}(2018)\citenamefont
  {Pezz\`e}, \citenamefont {Smerzi}, \citenamefont {Oberthaler}, \citenamefont
  {Schmied},\ and\ \citenamefont {Treutlein}}]{C-R1}%
  \BibitemOpen
  \bibfield  {author} {\bibinfo {author} {\bibfnamefont {L.}~\bibnamefont
  {Pezz\`e}}, \bibinfo {author} {\bibfnamefont {A.}~\bibnamefont {Smerzi}},
  \bibinfo {author} {\bibfnamefont {M.~K.}\ \bibnamefont {Oberthaler}},
  \bibinfo {author} {\bibfnamefont {R.}~\bibnamefont {Schmied}},\ and\ \bibinfo
  {author} {\bibfnamefont {P.}~\bibnamefont {Treutlein}},\ }\href
  {https://doi.org/10.1103/RevModPhys.90.035005} {\bibfield  {journal}
  {\bibinfo  {journal} {Rev. Mod. Phys.}\ }\textbf {\bibinfo {volume} {90}},\
  \bibinfo {pages} {035005} (\bibinfo {year} {2018})}\BibitemShut {NoStop}%
\bibitem [{\citenamefont {Giovannetti}\ \emph {et~al.}(2011)\citenamefont
  {Giovannetti}, \citenamefont {Lloyd},\ and\ \citenamefont {Maccone}}]{C-R2}%
  \BibitemOpen
  \bibfield  {author} {\bibinfo {author} {\bibfnamefont {V.}~\bibnamefont
  {Giovannetti}}, \bibinfo {author} {\bibfnamefont {S.}~\bibnamefont {Lloyd}},\
  and\ \bibinfo {author} {\bibfnamefont {L.}~\bibnamefont {Maccone}},\ }\href
  {https://doi.org/10.1038/nphoton.2011.35} {\bibfield  {journal} {\bibinfo
  {journal} {Nature Photonics}\ }\textbf {\bibinfo {volume} {5}},\ \bibinfo
  {pages} {222} (\bibinfo {year} {2011})}\BibitemShut {NoStop}%
\bibitem [{\citenamefont {C.Yuce}(2013)}]{Yuce_2013}%
  \BibitemOpen
  \bibfield  {author} {\bibinfo {author} {\bibnamefont {C.Yuce}},\ }\href
  {https://doi.org/10.1209/0295-5075/103/30011} {\bibfield  {journal} {\bibinfo
   {journal} {Europhysics Letters}\ }\textbf {\bibinfo {volume} {103}},\
  \bibinfo {pages} {30011} (\bibinfo {year} {2013})}\BibitemShut {NoStop}%
\bibitem [{\citenamefont {Rubio-Abadal}\ \emph {et~al.}(2020)\citenamefont
  {Rubio-Abadal}, \citenamefont {Ippoliti}, \citenamefont {Hollerith},
  \citenamefont {Wei}, \citenamefont {Rui}, \citenamefont {Sondhi},
  \citenamefont {Khemani}, \citenamefont {Gross},\ and\ \citenamefont
  {Bloch}}]{PhysRevX.10.021044}%
  \BibitemOpen
  \bibfield  {author} {\bibinfo {author} {\bibfnamefont {A.}~\bibnamefont
  {Rubio-Abadal}}, \bibinfo {author} {\bibfnamefont {M.}~\bibnamefont
  {Ippoliti}}, \bibinfo {author} {\bibfnamefont {S.}~\bibnamefont {Hollerith}},
  \bibinfo {author} {\bibfnamefont {D.}~\bibnamefont {Wei}}, \bibinfo {author}
  {\bibfnamefont {J.}~\bibnamefont {Rui}}, \bibinfo {author} {\bibfnamefont
  {S.~L.}\ \bibnamefont {Sondhi}}, \bibinfo {author} {\bibfnamefont
  {V.}~\bibnamefont {Khemani}}, \bibinfo {author} {\bibfnamefont
  {C.}~\bibnamefont {Gross}},\ and\ \bibinfo {author} {\bibfnamefont
  {I.}~\bibnamefont {Bloch}},\ }\href
  {https://doi.org/10.1103/PhysRevX.10.021044} {\bibfield  {journal} {\bibinfo
  {journal} {Phys. Rev. X}\ }\textbf {\bibinfo {volume} {10}},\ \bibinfo
  {pages} {021044} (\bibinfo {year} {2020})}\BibitemShut {NoStop}%
\end{thebibliography}%
\end{document}